\documentclass[]{aastex62}
\usepackage{longtable}
\usepackage{amsmath}

\begin{document}

\title{Bright $\gamma$-ray flares observed in GRB\,131108A}
\correspondingauthor{Donggeun Tak}
\author{M.~Ajello}
\affiliation{Department of Physics and Astronomy, Clemson University, Kinard Lab of Physics, Clemson, SC 29634-0978, USA}
\author{M.~Arimoto}
\affiliation{Faculty of Mathematics and Physics, Institute of Science and Engineering, Kanazawa University, Kakuma, Kanazawa, Ishikawa 920-1192}
\author{K.~Asano}
\affiliation{Institute for Cosmic-Ray Research, University of Tokyo, 5-1-5 Kashiwanoha, Kashiwa, Chiba, 277-8582, Japan}
\author{M.~Axelsson}
\affiliation{Department of Physics, Stockholm University, AlbaNova, SE-106 91 Stockholm, Sweden}
\affiliation{Department of Physics, KTH Royal Institute of Technology, AlbaNova, SE-106 91 Stockholm, Sweden}
\author{L.~Baldini}
\affiliation{Universit\`a di Pisa and Istituto Nazionale di Fisica Nucleare, Sezione di Pisa I-56127 Pisa, Italy}
\author{G.~Barbiellini}
\affiliation{Istituto Nazionale di Fisica Nucleare, Sezione di Trieste, I-34127 Trieste, Italy}
\affiliation{Dipartimento di Fisica, Universit\`a di Trieste, I-34127 Trieste, Italy}
\author{D.~Bastieri}
\affiliation{Istituto Nazionale di Fisica Nucleare, Sezione di Padova, I-35131 Padova, Italy}
\affiliation{Dipartimento di Fisica e Astronomia ``G. Galilei'', Universit\`a di Padova, I-35131 Padova, Italy}
\author{R.~Bellazzini}
\affiliation{Istituto Nazionale di Fisica Nucleare, Sezione di Pisa, I-56127 Pisa, Italy}
\author{A.~Berretta}
\affiliation{Dipartimento di Fisica, Universit\`a degli Studi di Perugia, I-06123 Perugia, Italy}
\author{E.~Bissaldi}
\affiliation{Dipartimento di Fisica ``M. Merlin" dell'Universit\`a e del Politecnico di Bari, I-70126 Bari, Italy}
\affiliation{Istituto Nazionale di Fisica Nucleare, Sezione di Bari, I-70126 Bari, Italy}
% ...
\begin{abstract}
GRB 131108A is a bright long Gamma-Ray Burst (GRB) detected by the Large Area Telescope and the Gamma-ray Burst Monitor on board the \textit{Fermi Gamma-ray Space Telescope}. Dedicated temporal and spectral analyses reveal three $\gamma$-ray flares dominating above 100 MeV, which are not directly related to the prompt emission in the GBM band (10 keV--10 MeV). The high-energy light curve of GRB 131108A (100 MeV -- 10 GeV) shows an unusual evolution: a steep decay, followed by three flares with an underlying emission, and then a long-lasting decay phase. The detailed analysis of the $\gamma$-ray flares finds that the three flares are 6 -- 20 times brighter than the underlying emission and are similar to each other. The fluence of each flare, (1.6 $\sim$ 2.0) $\times$ 10$^{-6}$ erg cm$^{-2}$, is comparable to that of emission during the steep decay phase, 1.7 $\times$ 10$^{-6}$ erg cm$^{-2}$. The total fluence from three $\gamma$-ray flares is 5.3 $\times$ 10$^{-6}$ erg cm$^{-2}$. The three $\gamma$-ray flares show properties similar to the usual X-ray flares that are sharp flux increases, occurring in $\sim$ 50\% of afterglows, in some cases well after the prompt emission. Also, the temporal and spectral indices during the early steep decay phase and the decaying phase of each flare show the consistency with a relation of the curvature effect ($\hat{\alpha}$ = 2 + $\hat{\beta}$), which is the first observational evidence of the high-latitude emission in the GeV energy band. 
\end{abstract}


\section{Introduction} \label{sec:intro}
Gamma-ray Bursts (GRBs), the most luminous electromagnetic events in the universe, show two emission phases: the prompt emission and the afterglow. The prompt emission, short and spiky pulses, dominates in the keV -- MeV energy range with multiple spectral components \citep[][and references therein]{Guiriec2015}. On the other hand, the light curve and spectrum of the afterglow, emission from the interaction between an outgoing blast wave from the central engine and a circumburst medium \citep{Meszaros1997, Sari1998}, are characterized by a series of broken power laws, sometimes accompanying bright flares \citep{Zhang2006, Nousek2006}. The afterglow is observed in a broad energy band from radio to $\gamma$-ray. The flares are commonly observed in the X-ray band \citep[e.g.,][]{Romano2006}, but rarely in the optical band \citep[e.g.,][]{Roming2006}. The X-ray flares have been explained as a result of the late-time activities of the central engine \citep[e.g.,][]{Fan2005, Zhang2006, Falcone2006, Falcone2007, Lazzati2007, Chincarini2007, Galli2007}.

Due to the curvature effect of a spherical, relativistic jet producing an X-ray flare, the decay phase of the X-ray flare evolves in a certain way. This effect was firstly discussed by \cite{Fenimore1996}, and \cite{Kumar2000} characterized the evolution of the temporal decaying index of the X-ray flare ($\hat{\alpha}$) as a function of corresponding spectral index ($\hat{\beta}$), $\hat{\alpha}$ = 2 + $\hat{\beta}$ in convention of $F_{\nu}$ $\propto$ $t^{-\hat{\alpha}}\nu^{-\hat{\beta}}$. This relation has been identified in many X-ray flares \citep[e.g.,][]{Liang2006, Chincarini2007, Uhm2016a, Jia2016}.

The \textit{Fermi Gamma-ray Space Telescope} (\textit{Fermi}) has observed numerous GRBs and helped to uncover the exotic high-energy evolution of GRBs. The high-energy emission ($>$ 100 MeV) of GRBs observed by the Large Area Telescope (LAT) on board \textit{Fermi} shares common features: delayed onset and lasting longer compared to keV -- MeV emission, requiring additional spectral components, and a power-law decaying light curve \citep{FLGC, 2FLGC}. These GeV features can be interpreted as the early afterglow emission \citep[e.g.,][]{Ghisellini2010, Kumar2010, Tak2019b}. \cite{Abdo2011} reported the GeV emission during vigorous X-ray flaring activities, but a flare above the underlying afterglow emission in the GeV energy band has not been reported before this work. 

In this work, we firstly report the three bright $\gamma$-ray flares observed in GRB\,131108A, which are $\sim$ 6 -- 20 times brighter than than underlying light curve (Figure~\ref{fig:LAT}). We will compare spectral and temporal properties of three $\gamma$-ray flares and the X-ray flares. The broadband spectral analysis and the correlation test between the low- and high-energy bands will be described.

\section{Observations}
\begin{figure}[t]
  \centering
  \includegraphics[scale=0.7]{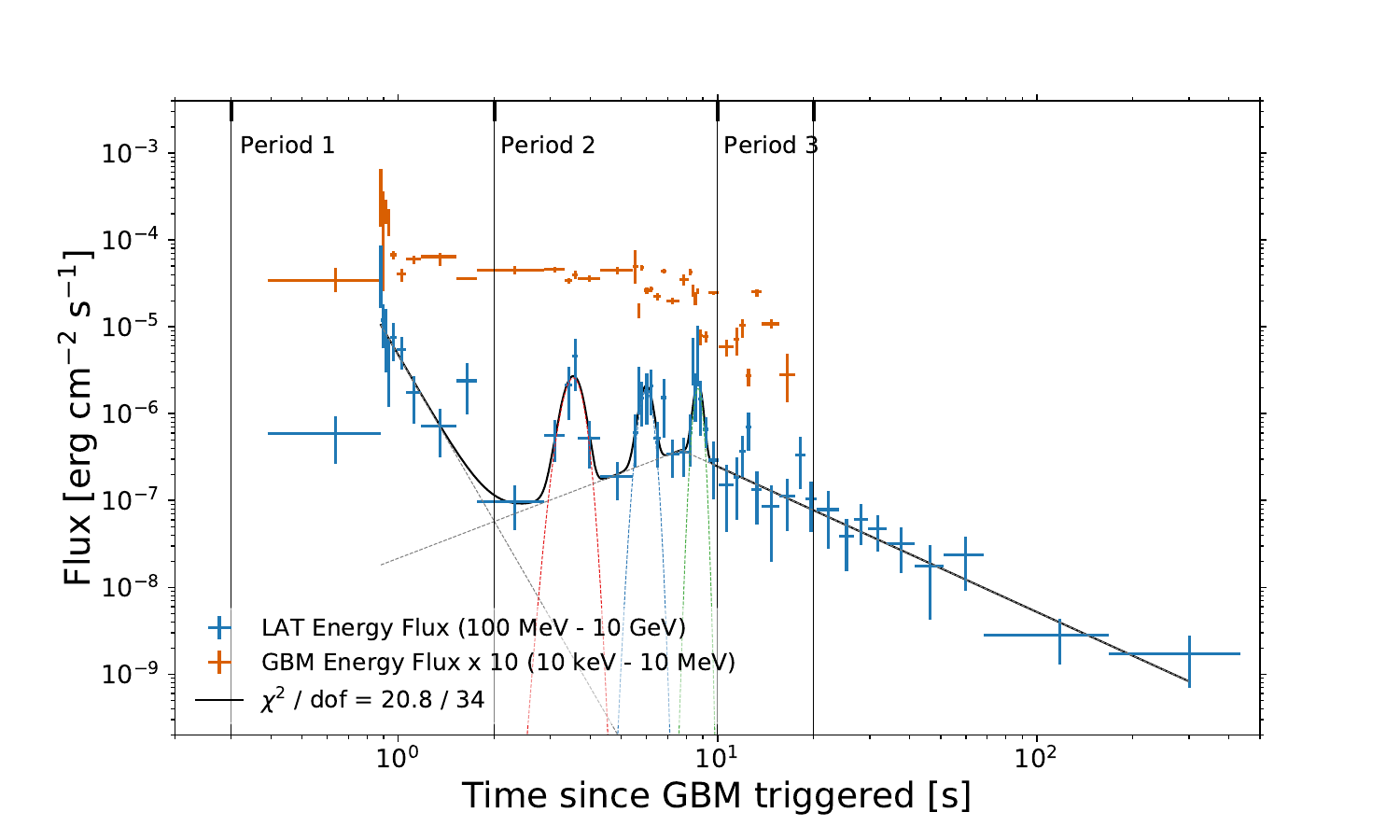}
  \caption{GBM and LAT light curves of GRB 131108A. The energy fluxes in the LAT energy band (100 MeV to 10 GeV) and in the GBM energy band (10 keV-- 1 MeV) are plotted in blue and orange, respectively. They are calculated from the best-fit model for each time interval in the spectral analysis with each instrument. The solid black line shows the fit of the LAT light curve consisting of five components: a simple power law (dotted gray line), a broken power law (dotted gray line), and three Gaussian functions (dotted red, green, and blue lines).}
  \label{fig:LAT}
\end{figure}

\begin{figure}[t]
  \centering
  \includegraphics[scale=0.8]{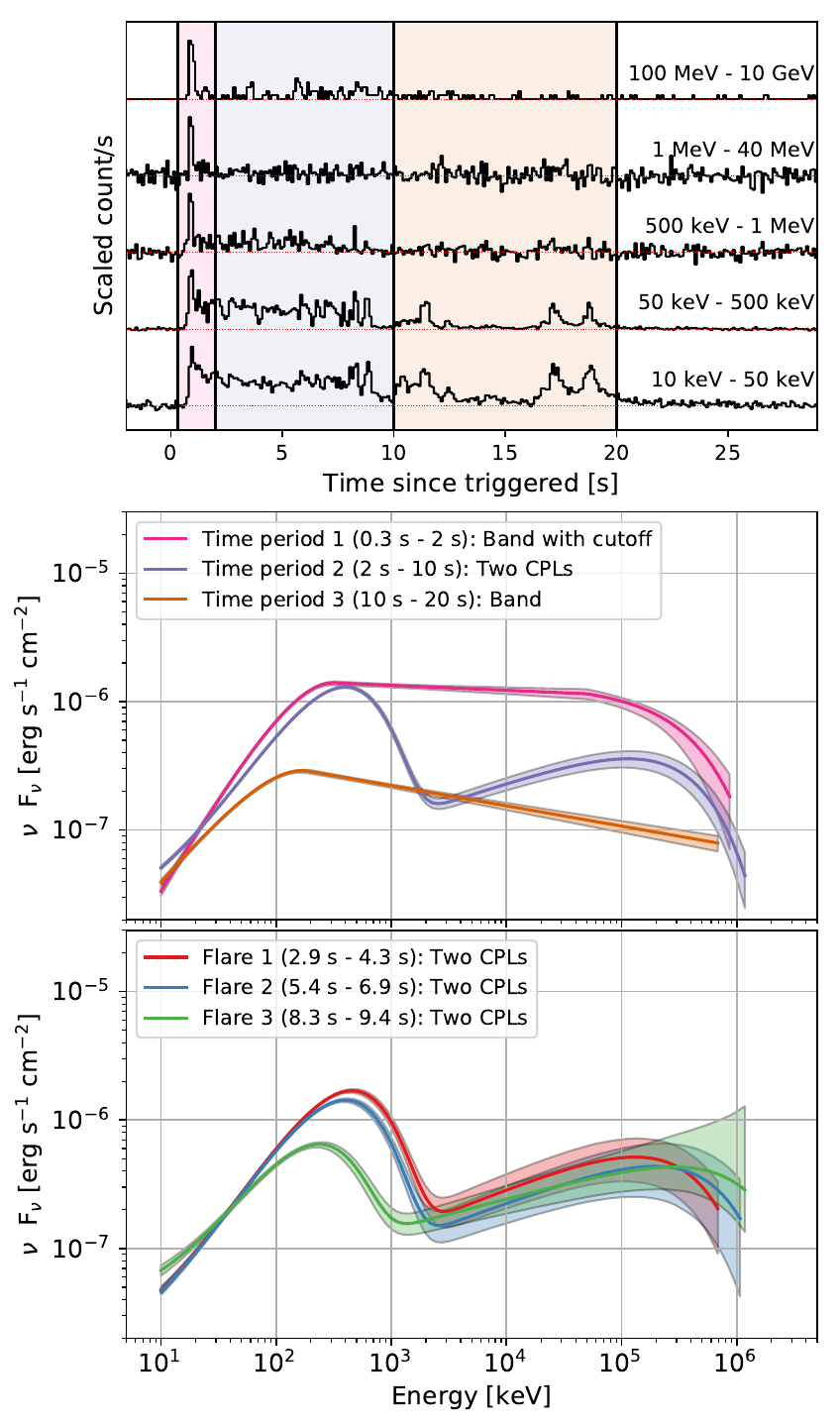}
  \caption{Count-rate curves and spectral energy distributions (SEDs). The top panel shows the scaled count-rate curves in different energy bands. The lower two panels show the joint-fit SEDs in energy band from 10 keV to 10 GeV. The color coding of the shaded region in the top panel and the spectrum in the middle panel indicates three time periods: pink (time period 1; 0.3 -- 2 s), violet (time period 2; 2 -- 10 s), and orange (time period3; 10 -- 20 s). The bottom panel shows SEDs for three GeV flares (red, blue, and green). Each solid curve represents the best-fit spectral shape (thick) with 1 confidence level contour (shaded region) derived from the errors on the fit parameters.} 
  \label{fig:SED}
\end{figure}

At 20:41:55.76 UTC on 2013 November 8 (T$_{0}$), LAT triggered on a bright high-energy emission from GRB 131108A \citep{GCN_LAT}, which is simultaneously observed by the Gamma-ray Burst Monitor (GBM) on board \textit{Fermi} \citep{GCN_GBM}. The duration of the burst (T$_{90}$ \footnote{a duration where a GRB emits from 5\% of its total counts to 95\%}), is 18.2 seconds, but the high-energy emission lasts $\sim$ T$_{0}$ + 600 seconds. With the observation of the bright afterglow of GRB 131108A by various instruments such as \textit{Swift} \citep{GCN_Swift, GCN_Swift2}, \textit{AGILE} \citep{GCN_Agile}, the accurate location and redshift of GRB 131108A were reported as (R.A., Dec.) = (156.50, 9.66) with an uncertainty of 3.6 arcsec in radius \citep{GCN_Swift2} and z $\sim$ 2.40 \citep{GCN_GTC}, respectively. \textit{Swift} XRT started to observe the afterglow of GRB 131108A 3.9 ks ($\sim$ 1 hour) after the GBM trigger, where the X-ray light curve decays smoothly in time \citep{GCN_Swift2}.

We perform a time-resolved analysis of LAT data in energy range of 100 MeV -- 10 GeV with the \textit{Fermi Science Tools (v11r5p3)}. We use \emph{``Transient020E''} class events with the standard cuts. Photons from within the 15-degree region of interest around GRB 131108A are considered, and the maximum zenith angle is set to 100 degrees. We fit a background rate from 3FGL sources \citep{3FGL2015}, the galactic diffuse emission (\emph{gll\_iem\_v06}), and the isotropic diffuse emission (\emph{iso\_P8R2\_TRANSIENT020\_V6\_v06})\footnote{\url{https://fermi.gsfc.nasa.gov/ssc/data/access/lat/BackgroundModels.html}}. The LAT events observed in GRB 131108A are binned. For four sequential LAT events, we perform an unbinned likelihood analysis, and compute a test statistic (TS)\footnote{the detection significance of the source above the background} for the burst. If the resultant TS is lower than 9 (equivalent to 3 $\sigma$), we add the next event to the bin and compute the TS again. Once we have the bin with the TS $\geq$ 9, we collect the following four events, and repeat this procedure. As a result, each bin contains at least four LAT events, resulting in a TS $\geq$ 9. For each of these bins, we perform an unbinned maximum likelihood fit on the energy spectrum with a simple power-law (PL) model.

The high-energy light curve of GRB 131108A shows an unusual evolution compared to other bright LAT GRBs  (Figure~\ref{fig:LAT}); rather, it resembles the canonical X-ray early afterglow light curve though compressed to earlier and shorter timescales (seconds to tens of seconds compared to hundreds to thousands of seconds) \citep{Zhang2006, Nousek2006}. We find the best description of the LAT light curve by fitting several models and their parameters with the maximum likelihood method. The light curve is well-fitted with five components ($\chi^{2}$ / dof = 20.8 / 34): a simple power law, a broken power law, and three Gaussian functions for the three bright pulses above an underlying emission. Note that single-component models such as a simple power law ($\chi^{2}$ / dof = 130.8 / 47) or a broken power law ($\chi^{2}$ / dof = 90.8 / 45) are not a good model for this light curve. The model of each pulse can be replaced with the Norris function \citep{Norris1996} ($\chi^{2}$ / dof = 20.5 / 31), a broken power law ($\chi^{2}$ / dof = 20.2 / 31), or a smoothly broken power law \citep{Liang2006} ($\chi^{2}$ / dof = 19.4 / 28). However, the Gaussian function is the best-fit model considering its statistics and the number of free parameters. The best-fit parameters for the three Gaussian functions are listed in Table~\ref{tab:flare}. Note that there are hints of more than three flares but other fluctuations are insignificant, which are composed of one or two flux point. The fluence of each pulse is (2.0 $\pm$ 0.8), (1.6 $\pm$ 0.6), and (1.7 $\pm$ 0.7) $\times$ 10$^{-6}$ erg cm$^{-2}$, totally (5.3 $\pm$ 1.2) $\times$ 10$^{-6}$ erg cm$^{-2}$. The fluence of each pulse is comparable to that of emission during the early steep decay phase, (1.7 $\pm$ 0.4) $\times$ 10$^{-6}$ erg cm$^{-2}$. The decaying index of the later segment of the broken power law is 1.6 $\pm$ 0.2, consistent with other Fermi-LAT GRBs \citep{2FLGC}. 

Considering the LAT light curve and its best-fit model, we define three time periods (Figure~\ref{fig:SED}): the early steep decay period (time period 1; T$_{0}$+0.3 s -- T$_{0}$+2 s), the three $\gamma$-ray unusual pulses with the underlying emission period (time period 2; T$_{0}$+2 s -- T$_{0}$+10 s), and the long-lasting shallow decay period (time period 3; T$_{0}$+10 s -- T$_{0}$+20 s). Note that the third time period can be extended until the end of the LAT emission, but stops at the end of the prompt emission for the joint-fit purpose. The evolution of the first and last periods is commonly seen in the LAT GRBs, but the phenomena of the second time period are noteworthy.

For the three time periods, we perform a broadband spectral analysis with GBM and LAT data in energy band from 10 keV to 10 GeV. Of the twelve NaI and two BGO detectors that make up GBM, four NaI detectors (0, 3, 6, and 7; 10 keV -- 1 MeV) and two BGO detectors (0 and 1; 200 keV -- 40 MeV) show a considerably high count rate above the background level, so that we constitute a set of data from these detectors. In addition to the GBM data, LAT Low Energy (LLE; 30 MeV -- 100 MeV) and LAT (\emph{Transient020E}; 100 MeV -- 10 GeV) data are used. The background rate of GBM and LLE data is estimated by making use of the analysis package, \textit{RMfit (version43pr2)}, by fitting a time interval combined before and after the prompt emission phase of the burst with a polynomial function. The LAT background is estimated with ``gtbkg''\footnote{This tool generates a background spectrum file, which contains the total background rate from 3FGL sources, the galactic diffuse source, and the isotropic diffuse source.} provided by the \textit{Fermi Science Tools}. The energy spectrum of each time period is fitted with various models by using the maximum likelihood method with \textit{Xspec (12.9.1)} \citep{Arnaud1996}. We use a Poisson data with Gaussian background statistic (PG-stat) for the parameter estimation, and then use the Bayesian Information Criterion \citep[BIC;][]{Schwarz1978} for comparing the likelihood of fit and selecting the best-fit model. The best-fit model is a model with the lowest BIC value.

In the first time period characterized by the short bright emission commonly observed in the broad energy band from 10 keV to 10 GeV, the best-fit model for this time period is the Band function (Band) \citep{Band1993} with a high-energy cutoff\footnote{
\begin{equation}
\frac{dN}{dE} =
\begin{cases}
N_{0}\left(\frac{E}{100 \rm keV}\right)^{\alpha}exp\left(-\frac{E(\alpha+2)}{E_{p}}\right) &\text{if }E \leq \frac{\alpha-\beta}{2+\alpha}E_p,\\
N_{0}\left(\frac{E}{100 \rm keV}\right)^{\beta}(\frac{E_p}{100 \rm keV}\frac{\alpha-\beta}{2+\alpha})^{\alpha-\beta}exp(\beta-\alpha) &\text{if }\frac{\alpha-\beta}{2+\alpha}E_p < E \leq E_c,\\
N_{0}\left(\frac{E}{100 \rm keV}\right)^{\beta}(\frac{E_p}{100 \rm keV}\frac{\alpha-\beta}{2+\alpha})^{\alpha-\beta}exp(\beta-\alpha)exp\left(\frac{E_c-E}{E_{f}}\right) & \text{if }E > E_c,
\end{cases}
\end{equation}
where $\alpha$ and $\beta$ are the low- and high-energy photon indices, respectively, E$_{p}$ is the peak energy of the Band function, E$_{c}$ is the cutoff energy which is fixed to 50 MeV, and E$_{f}$ is the e-folding energy for the high-energy cutoff} (Table~\ref{tab:result}). The decrease in BIC as a result of adding the high-energy cutoff to the Band function is $\sim$ 17 units, implying that the high-energy cutoff is strongly required. One alternative model is a combination of two spectral components, Band and a power-law with an exponential cutoff (CPL)\footnote{$\frac{dN}{dE} = N_{0}\left(\frac{E}{100 keV}\right)^{\alpha}exp\left(-\frac{E(\alpha+2)}{E_p}\right)$}. This Band + CPL model describes the data slightly better (lower PG-stat), but the statistical improvement is not high enough to compensate the increase of a free parameter, making BIC higher than the best-fit model. 

The second time interval where we found the unusual pulses shows a high count rate only in the low- (10 keV--1 MeV) and high-energy regimes (100 MeV--10 GeV) (upper panel in Figure~\ref{fig:SED}). The observed data is best explained by a two-component model, CPL + CPL\footnote{$\frac{dN}{dE} = N_{0, \rm low}\left(\frac{E}{100 keV}\right)^{\alpha}exp\left(-\frac{E(\alpha+2)}{E_{p, \rm low}}\right) + N_{0, \rm high}\left(\frac{E}{100 keV}\right)^{-\Gamma}exp\left(-\frac{E(-\Gamma+2)}{E_{p, \rm high}}\right)$} (orange in Figure~\ref{fig:SED}, which is preferred over a single-component model such as Band (Table~\ref{tab:result}). The two CPLs have distinct peak energies, E$_{p, \rm low}$ $\sim$ 400 keV and E$_{p, \rm high}$ $\sim$ 130 MeV, respectively (Table~\ref{tab:result}). When any one of the CPL components is replaced with the Band function, $\beta$ becomes very soft so that the high-energy segment of Band is indistinguishable from the exponential cutoff. Therefore, the combination of Band and CPL is not necessary. The two-component scenario for GRB 131108A is also reported by \cite{Giuliani2014}, who analyzed the \textit{AGILE} (350 keV--30 GeV) data and reached the conclusion that the extrapolation of the low-energy spectral component could not explain the high-energy emission, and an additional spectral component with a peak energy at few MeV is required. The CPL dominating in the low-energy band has $\alpha$ $\simeq$ -0.5 consistent with that of the best-fit model in the first time period (Table~\ref{tab:result}), implying that the low-energy emission of the first and second time periods may be continuous. Given the best-fit model, the LAT emission is described by the high-energy CPL component. In addition, we perform a time-resolved spectral analysis for time intervals during the three high-energy pulses, and two distinct spectral components are again observed (Table~\ref{tab:result} and Figure~\ref{fig:SED}). The fluence of this high-energy component during the second time period is 5.9$\substack{+0.5\\-0.8 }$ $\times$ 10$^{-6}$ erg cm$^{-2}$, comparable to the sum of fluence from three $\gamma$-ray pulses, (5.2 $\pm$ 1.2) $\times$ 10$^{-6}$ erg cm$^{-2}$ (Table~\ref{tab:flare}). The most of the LAT emission during the second time period can be dominated by the three $\gamma$-ray pulses, and thus the high-energy CPL component may represent the spectral shape of the three $\gamma$-ray pulses. 

During the third time period, short-soft pulses in the low-energy band ($<$ 500 keV) are observed. The best-fit model in this time period is the Band function (green in Figure~\ref{fig:SED}). A CPL + CPL model does not give a better result, which requires the two more parameters but resulting in the similar statistics (Table~\ref{tab:result}). After T$_{0}$ + 20 s, the LAT spectrum is well-described by a power law with a photon index $\Gamma$ = 2.8 $\pm$ 0.3.

Figure~\ref{fig:LAT} and the upper panel of Figure~\ref{fig:SED} show that the low- (keV to few MeV) and high-energy (100 MeV--10 GeV) light curves evolve differently, and the broadband spectral analysis reveals the presence of the two spectral components. We check correlation between the low- and high-energy light curves with the Discrete Correlation Function \citep{Edelson1988}, which compares the variability of two light curves and estimates the time lag and the respective cross-correlation coefficient \citep[e.g.,][]{Rani2009}. For this purpose, we performed a time-resolved spectral analysis for the time bin used for the LAT light curve (Figure~\ref{fig:LAT}) with the GBM data, and computed flux in the range of 10 keV--10 MeV with the best-fit model for each time interval (orange points in Figure~\ref{fig:LAT}). When the entire interval is considered, the correlation between the two light curves, 10 keV--10 MeV and 100 MeV--10 GeV, is evident (correlation coefficient peak = 0.8 $\pm$ 0.1). However, if only the light curves in the second time period is considered, the correlation analysis does not suggest any correlation between the two light curve (correlation coefficient peak $\sim$ 0.1).

Considering the temporal and spectral features, the $\gamma$-ray pulses invoke a distinct origin from the prompt emission of the low-energy band as well as the LAT extended emission. From now on, the individual $\gamma$-ray pulse is called a ``$\gamma$-ray flare''. 

\section{Discussion}

\begin{figure*}[t]
  \centering
  \includegraphics[scale=0.48, trim=100 0 100 0, clip]{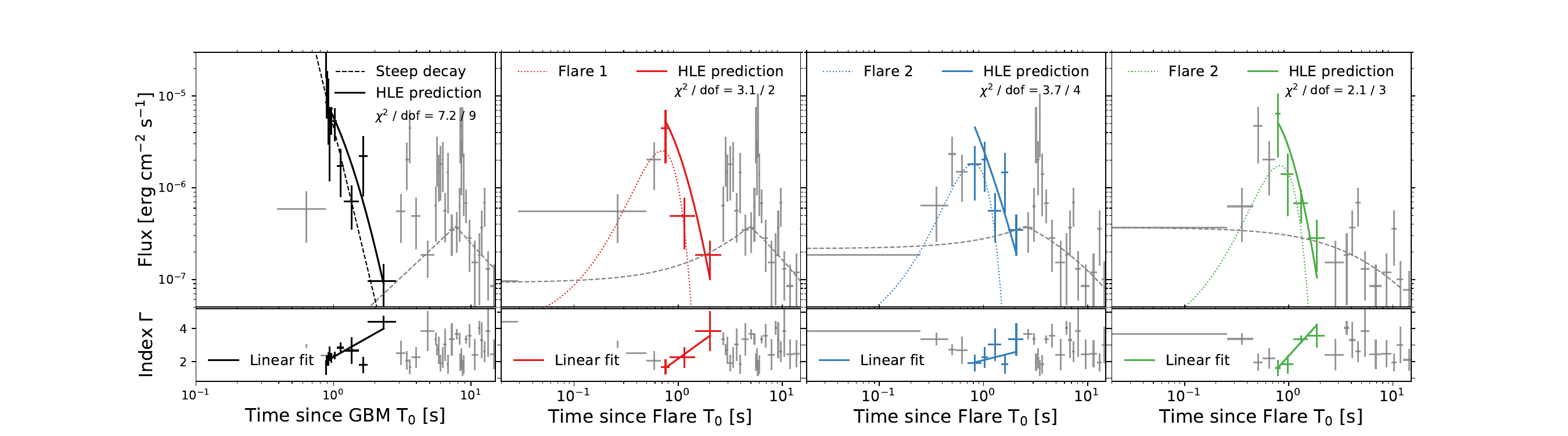}
  \caption{Test of the curvature effect for early steep decay emission and three $\gamma$-ray flares. The upper panels show the light curve of the early steep decay emission and the three $\gamma$-ray flares separately after removing the t$_{0}$ effect. The lower panel shows the evolution of the photon index. The data points corresponding to the decay phase of each flare are in red, blue, and green. The early steep decay phase is marked in black. The solid lines in the upper panels show the expected light curve derived from the relation of the curvature effect, $\hat{\alpha}$ = 2 + $\hat{\beta}$. These theoretical light curves are computed from the linear fit of the photon indices of the decay phase (solid line in the lower panel). }
  \label{fig:HLE}
\end{figure*}

First of all, we compare and test the well-known properties of X-ray flares to the observation of the $\gamma$-ray flares in GRB 131108A. 

A flux variation of the X-ray flares, a flux ratio of a flare to an underlying emission ($\delta F_{\nu}/F_{\nu}$), ranges from 6 \citep[e.g. GRB 050406;][]{Romano2006} to higher than 100 \citep[e.g., GRB 050202B;][]{Burrows2005}. The $\gamma$-ray flares are about 6 to 20 times brighter than the underlying emission, which slowly changes in time, $F_{\nu}$ $\sim$ 2.7 $\times$ 10$^{-7}$ erg cm$^{-2}$ s$^{-1}$ on average (see Figure~\ref{fig:LAT}), and thus the flux variation of the $\gamma$-ray flares is sub-normal, compared to X-ray flares \citep{Chincarini2007}. The duration of the X-ray flares varies from few hours to a day \citep{Chincarini2007, Swenson2014}, and there is an empirical relation between the onset time and the duration of the X-ray flares, $\delta$t/t $\sim$ 0.1 \citep[e.g.,][]{Chincarini2007, Chincarini2010, Swenson2014}. In case of the $\gamma$-ray flares, they last only few seconds (Table~\ref{tab:flare}), much shorter than the X-ray flares \citep{Chincarini2007}. Also, the $\gamma$-ray flares are observed in $<$ T$_{0}$ + 10 s, which is
earlier than any X-ray flares \citep{Chincarini2007}. 
Combining these two unusual features, surprisingly, the temporal characteristics of the $\gamma$-ray flares are not in conflict with the empirical relation. A comparison between the flux variability and the temporal variability of the $\gamma$-ray flares ($\delta F_{\nu}/F_{\nu}$ vs. $\delta$t/t) shows that the $\gamma$-ray flares are consistent with X-ray flares \citep{Chincarini2007}. Furthermore, this comparison implies that the $\gamma$-ray flares are not related to the fluctuations of the external shock as previously discussed for the X-ray flares \citep{Ioka2005, Zhang2006}.

The steep decay of the X-ray flares is regarded as a result of the curvature effect, which is identified by testing the relation, $\hat{\alpha}$ = 2 + $\hat{\beta}$ \citep[e.g.,][]{Liang2006, Chincarini2007, Uhm2016a, Jia2016}. It is possible that the decay phase of the $\gamma$-ray flares also show evidence of the curvature effect. Before testing the relation, we should remove the so-called t$_{0}$ effect \citep{Zhang2006, Kobayashi2007}. Each flare is attributed to the late-time activity of the central engine and thus has its own onset time (t$_{0}$). Since the shape of a light curve in the logarithmic space is very sensitive to the choice of t$_{0}$, the intrinsic light curve of the flare can only be provided if the light curve is shifted to the true t$_{0}$. Due to the underlying emission, however, the true onset of the $\gamma$-ray flares is ambiguous. Therefore, we properly choose the onset of each flare as the time when the flux of the flare is 1/100 of its peak. Figure~\ref{fig:HLE} shows the $\gamma$-ray flares after shifting them to the proper t$_{0}$ for each flare. For these light curves, we test the curvature effect relation. After selecting the data points corresponding to the decaying phase, we fit the measured photon indices ($\Gamma$) with a linear function, $\Gamma$ = $f(t-t_{0})$ (solid line in lower panel of Figure~\ref{fig:HLE}). Next, the photon index is converted to the spectral index, $\hat{\beta}$ = $\Gamma$ - 1. We then apply the HLE relation and get the temporal index as a function of time, $\hat{\alpha}$ = $f(t-t_{0})$ + 1. Finally, the light curve expected by the curvature effect is described by a function of time, $F_{\nu}$ = $F_{\nu,0}$ $(t-t_{0})^{f(t-t_{0}) + 1}$ (solid line in upper panel of Figure~\ref{fig:HLE}). We fit this function with the observed, shifted flux points and conclude that the decay phases of all three $\gamma$-ray flares are consistent with the expectation by the curvature effect (Figure~\ref{fig:HLE}). Also, we find the spectral softening during the decay phase of the flares, which is the well-known phenomenon identified in the X-ray flares \citep{Chincarini2007, Falcone2007}. 

The X-ray flares are likely attributed to internal shocks, where accelerated electrons at the shocks radiate via the synchrotron process. On the other hand, the $\gamma$-ray flares with E$_{p}$ $\sim$ 130 MeV may originate from the Synchrotron Self-Compton (SSC) process from same population of electrons that might have produced X-ray flares. In principle, there could be two possible cases for the inverse Compton process: SSC from the internal emission region and the External Inverse Compton (EIC) from the external shock region \citep[][]{Wang2006, Fan2008}. The observation of the high-latitude emission in the flares disfavors the EIC origin and supports the SSC origin. Assuming the typical electron Lorentz factor, $\gamma_{e}$ $\gtrsim$ 10$^{3}$, the peak energy of the seed photon should be $\lesssim$ 0.13 keV, which is far below the GBM energy band. Another possibility for the origin of the $\gamma$-ray flares is the residual collision in the internal dissipation process \citep{Li2008}. In this case, the $\gamma$-ray flares can be interpreted as the SSC counterpart of the optical emission produced by the residual collision at large radii. Note that there were no X-ray and optical observations during the prompt emission phase of this burst, so that these hypotheses cannot be tested.

The very first steep decay emission in the first time period corresponds to the tail of the first bright broadband pulse. This decay emission is also consistent with the curvature effect (the first panel in Figure~\ref{fig:HLE}). 

The underlying emission in the second time period can be interpreted as the emission during the development of the forward shock \citep[e.g.,][]{Maxham2011}, and the long-lasting decay emission (the third time period) can be the continuous emission from the fully developed forward shock when the total energy is not noticeably increased by the additional energy injection \citep[e.g.,][]{Meszaros1997, Sari1998}. 

The observation of GRB 131108A uncovers a new phenomenon in the high-energy GRB light curve. Even though the three $\gamma$-ray flares were observed in the prompt phase of the burst, they showed the temporal and spectral properties similar to those of the X-ray flares. Also, we found the evidence of the curvature effect in the GeV energy band for the first time. 
\begin{table*}
\small 
\centering 
\caption{The physical properties of three flares}
	\begin{tabular}{c c c c c}
    \hline\hline
    &Peak flux\footnote{in 100 MeV -- 10 GeV} & Peak time & FWHM \footnote{Full width at half maximum} & Fluence$^{a}$ \\
    & [\,$\times$ 10$^{-6}$ erg cm$^{-2}$ s$^{-1}$\,] & [\,s\,] & [\,s\,] & [\,$\times$ 10$^{-6}$ erg cm$^{-2}$\,] \\\hline
    Flare 1 & 2.6 $\pm$ 1.3 & 3.5 $\pm$ 0.1 & 0.8 $\pm$ 0.2 & 2.0 $\pm$ 0.9\\ 
    Flare 2 & 1.4 $\pm$ 0.6 & 6.0 $\pm$ 0.1 & 0.9 $\pm$ 0.2 & 1.6 $\pm$ 0.6 \\
    Flare 3 & 2.4 $\pm$ 1.0 & 8.5 $\pm$ 0.1 & 0.9 $\pm$ 0.2 & 1.6 $\pm$ 0.6 \\\hline\hline
\end{tabular}

\label{tab:flare}
\end{table*}

\begin{table*}
\small 
\centering 
\caption{Joint-fit spectral analysis parameters.}
\begin{tabular}{c | c | c c c c c c | c c c}
\hline\hline
Time period & Model & $\alpha$ & $\beta$ & $\Gamma$ & E$_{p, \rm low}$ & E$_{p, \rm high}$ & E$_{f}$ & PG-stat & dof & BIC \\
 & & & & & [ keV ] & [ MeV ]  & [ MeV ] &  &  &  \\\hline
1 & Band & -0.50$^{+0.06}_{-0.06}$ & -2.07$^{+0.02}_{-0.02}$ & & 320.6$^{+22.3}_{-20.1}$ & & & 745 & 695 & 771 \\ 
(0.3 -- 2 s)& Band with highcut\footnote{The cutoff energy E$_{c}$ is fixed to 50 MeV} & -0.48$^{+0.07}_{-0.06}$ & -2.01$^{+0.02}_{-0.02}$ & & 305.3$^{+21.5}_{-19.8}$ && 350.3$^{+143.1}_{-87.6}$ & 722 & 694 & 754\\
\hline
2 & Band & -0.73$^{+0.02}_{-0.02}$ & -2.24$^{+0.01}_{-0.01}$ & & 389.7$^{+14.2}_{-13.4}$ & & & 880 & 695 & 906\\ 
(2 -- 10 s) & CPL + CPL & -0.50$^{+0.07}_{-0.07}$ &  & 1.69$^{+0.02}_{-0.03}$& 394.2$^{+13.3}_{-12.4}$ &126.2$^{+25.6}_{-19.5}$  && 821 & 693 & 860\\\hline
3 & Band & -0.94$^{+0.05}_{-0.05}$ & -2.15$^{+0.02}_{-0.02}$ & & 168.0$^{+12.5}_{-11.2}$ & & &  741 & 695 & 767\\
(10 -- 20 s) & CPL + CPL & -0.92$^{+0.13}_{-0.09}$ & &  1.68$^{+0.07}_{-0.11} $ & 198.9$^{+13.7}_{-12.0}$ & 68.7$^{+21.5}_{-15.5}$ && 742 & 693 & 781\\ \hline
Flare 1 & CPL + CPL & -0.5$_{\rm fixed}$\footnote{The photon index for the low-energy CPL component is fixed to -0.5, which is the photon index of the best-fit model in the second time interval.} &  & 1.64$^{+0.04}_{-0.04}$& 447.9$^{+19.9}_{-18.7}$ &131.3$^{+63.3}_{-35.7}$  && 819 & 692\footnote{Note that the change in the dof results from decrease in the number of the energy bin of LAT data.} & 852\\
Flare 2 & CPL + CPL & -0.5$_{\rm fixed}$$^{\rm b}$ &  & 1.67$^{+0.04}_{-0.04}$& 396.0$^{+18.1}_{-16.9}$ &193.1$^{+126.7}_{-62.0}$  && 763 & 692$^{\rm c}$ & 796\\
Flare 3 & CPL + CPL & -0.5$_{\rm fixed}$$^{\rm b}$ &  & 1.76$^{+0.03}_{-0.03}$& 229.0$^{+17.5}_{-15.7}$ &284.9$^{+419.4}_{-124.4}$  && 716 & 692$^{\rm c}$ & 749\\
\hline\hline
\end{tabular}
\label{tab:result}
\end{table*}

\acknowledgments
\section*{Acknowledgement}
The \textit{Fermi}-LAT Collaboration acknowledges support for LAT development, operation and data analysis from NASA and DOE (United States), CEA/Irfu and IN2P3/CNRS (France), ASI and INFN (Italy), MEXT, KEK, and JAXA (Japan), and the K.A.~Wallenberg Foundation, the Swedish Research Council and the National Space Board (Sweden). Science analysis support in the operations phase from INAF (Italy) and CNES (France) is also gratefully acknowledged. This work performed in part under DOE Contract DE-AC02-76SF00515.
\software{XSPEC \citep[v12.9.1;][]{Arnaud1996}, Fermi Science Tools (v11r5p3)}
\bibliographystyle{aasjournal}
\bibliography{references}

\end{document}